\documentclass[aps,prl,twocolumn,groupedaddress,showpacs]{revtex4}
\usepackage{graphicx,amsmath,amssymb,amsfonts,latexsym,color,dcolumn,bm}
\usepackage{verbatim}

\providecommand{\ket}[1]{\lvert #1 \rangle}

\begin{document}

\title{Coherent frequency conversion in a superconducting artificial atom with two internal degrees of freedom}

\author{F. Lecocq, I. M. Pop, I. Matei, E. Dumur, A. K. Feofanov, C. Naud, W. Guichard, O. Buisson}

\affiliation{Institut N\'eel, C.N.R.S.- Universit\'e
Joseph Fourier, BP 166, 38042 Grenoble-cedex 9, France}

\date{\today}

\begin{abstract}

By adding a large inductance in a dc-SQUID phase qubit loop, one decouples the junctions' dynamics and creates a superconducting artificial 
atom with two internal degrees of freedom. In
addition to the usual symmetric plasma mode ({\it s}-mode) which gives rise to the phase qubit, an
anti-symmetric mode ({\it a}-mode) appears. These two
modes can be described by two anharmonic oscillators with
eigenstates $\ket{n_{s}}$ and $\ket{n_{a}}$ for the {\it s} and {\it
a}-mode, respectively. We show that a strong
nonlinear coupling between the modes leads to a large energy
splitting between states $\ket{0_{s},1_{a}}$ and
$\ket{2_{s},0_{a}}$. Finally, coherent frequency conversion is observed via free oscillations between the states
$\ket{0_{s},1_{a}}$ and $\ket{2_{s},0_{a}}$.

\end{abstract}

\maketitle

In atomic physics the presence of multiple degrees of freedom (DoF)
 constitutes a precious resource for the development of quantum
mechanics experiments. In trapped ions or nitrogen-vacancy centers, the V-shaped
energy spectrum enables very high fidelity readout of the states
encoded in the first two levels by using the fluorescence properties
of the transition to the third level \cite{Leibfried_Review2003,Jelezko_PRL2004}.
 In quantum optics, the multiple DoF of the optically active crystals have lead to many quantum effects 
 such as Coherent Population Trapping (CPT) and
the associated effect of Electromagnetically Induced Transparency
(EIT) \cite{Fleischhauer_Review2003}, spontaneous emission
cancelation via quantum interference \cite{Zhu_PRL1996,Xia_PRL1996} or generation of entangled photon pairs \cite{Ou_PRL1988}.
 In the field of superconducting qubits, experimental efforts have
mainly focused on two-level systems
\cite{Korotkov_09,Clarke_2008} and multilevel systems \cite{Claudon_PRL2004,Lucero_PRL2008,Dutta_PRB2008,Bishop_NatPhys2008,Bianchetti_PRL2010} with a single DoF. Superconducting artificial atoms currently need additionnal DoF in order to realize
$\Lambda$, $V$, $N$ or diamond-shaped energy levels and therefore to perform new quantum experiments \cite{You_Nature2011,Hu_PRA2011}. Only recently a superconducting device with $V$-shape energy spectrum was experimentally considered
\cite{Srinivasan_PRL2011} and a three DoF superconducting ring was developed for parametric amplification \cite{Bergeal_NatPhys2010}. 

In this letter we present an artificial atom with two internal DoF which constitutes a basic block for the realization of V- or Diamond-shaped energy levels. In this system we benefit from the natural nonlinear coupling between the two DoF to observe  a coherent frequency conversion process in the time domain. Contrary to previous frequency conversion proposals and implementations in solid state devices, we observe this process in the strong coupling limit, where multiples oscillations can be seen before losing coherence \cite{Moon_PRL2005,Marquardt_PRB2007}, and without any external coupling device or additionnal source of power \cite{EvaZakka_NatPhys2011}. In addition this system could be used for triggered and high-efficiency generation of entangled pairs of photons, a key component for quantum information \cite{Bouwmeester_2000}.

\begin{figure}[htb*]
\includegraphics[bb=0bp 380bp 245bp 694bp,clip,width=0.48\textwidth]{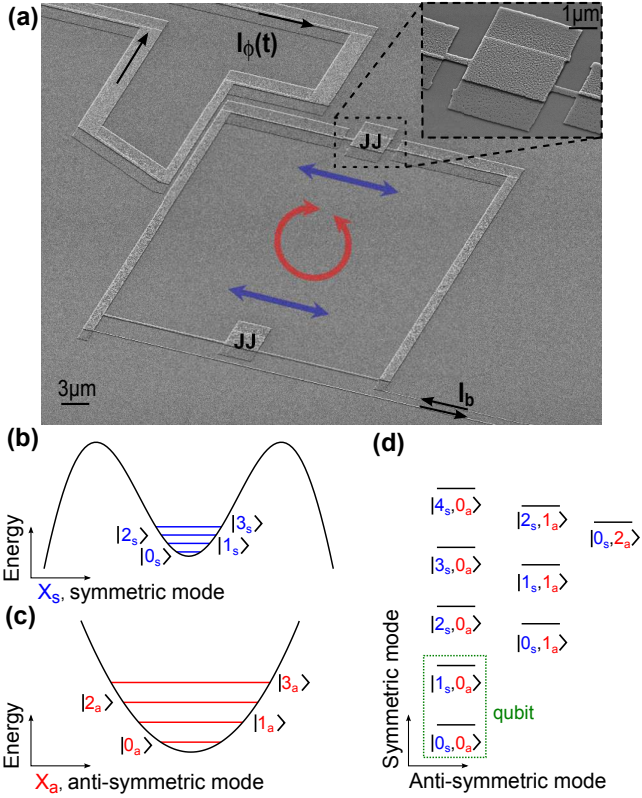}
\caption{\textbf{Description of the device}. \textbf{(a)} A
micrograph of the aluminium circuit. The two small squares are the
two Josephson junctions (enlarged in the top right inset, $10\mu
m^2$ area, $I_c=713$~nA and $C = 510$~fF) decoupled by a large inductive loop
($L=629$~pH). The width of the two SQUID arms were adjusted to reduce
the inductance asymmetry to about 10\%. Very narrow current bias
lines, with a large $15$~nH-inductance, isolate the
quantum circuit from the dissipative environment at high frequencies . The symmetric and
antisymmetric oscillation modes are illustrated by blue and red
arrows, respectively. \textbf{(b)} and \textbf{(c)} Potentials of the
{\it s} and {\it a}-mode respectively, for the bias working point
($I_b=0$,$\Phi_b=0.37\Phi_0$). \textbf{(d)} Schematic energy level
diagram indexed by the quantum excitation number of the two modes
$\ket{n_{s},n_{a}}$ at the same working point. Climbing each
vertical ladder one increases the excitation number of the {\it
s}-mode, keeping the excitation number of the {\it a}-mode constant.
The two first levels, $\ket{0_{s},0_{a}}$ and $\ket{1_{s},0_{a}}$,
realize a camelback phase qubit.\label{fig1}}
\end{figure}

The circuit is a camelback phase qubit \cite{Hoskinson_PRL2009} with a large loop inductance, {\it i.e} a dc-SQUID build by a superconducting loop of large inductance $L$ interrupted by two identical Josephson junctions with critical current $I_{c}$
and capacitance $C$, operated at zero current bias (see Fig.\ref{fig1}). As we will see in the following, the presence of a large loop inductance modifies dractically the quantum dynamics of this system.
The two phase differences $\phi_1$ and $\phi_2$
across the two junctions correspond to the two degrees of freedom of
this circuit, which lead to two oscillating modes: the symmetric ({\it
s}-) and the anti-symmetric ({\it a}-) plasma modes \cite{footnote1}. The
{\it s}-mode corresponds to the well-known in-phase plasma
oscillation of the two junctions with the average phase
$x_{s}=(\phi_1+\phi_2)/2$, oscillating at a characteristic frequency
given by the plasma frequency of the dc-SQUID, $\omega_p^s$. The {\it
a}-mode is an opposite-phase plasma mode related to oscillations of
the phase difference $x_{a}=(\phi_1-\phi_2)/2$, producing
circulating current oscillations at frequency $\omega_p^a$. In
previous experiments
\cite{Claudon_PRL2004,Dutta_PRB2008,Hoskinson_PRL2009} , the loop
inductance $L$ was small compared to the Josephson inductance $L_J=\Phi_0/(2\pi I_{c})$.
 Therefore the two junctions were
strongly coupled and the dynamics of the phase difference $x_{a}$
was neglected and fixed by the applied flux. The quantum behavior
of the circuit was described by the {\it s}-mode  only, showing a
one-dimensional motion of the average phase $x_{s}$. Hereafter we
will consider a circuit with a large inductance ($L \geq L_J$) that
decouples the phase dynamics of the two junctions. This large
inductance lowers the frequency of the {\it a}-mode and the
dynamics of the system becomes fully two-dimensional. The {\it a}-mode was
previously introduced to discuss the thermal and quantum escape of a
current-biased dc-SQUID \cite{Lefevre_PRB1992,Balestro_PRL2003} but
its dynamics was never observed. We present measurements of the full
spectrum of this artificial atom, independent coherent control of
both modes and finally we exploit the strong nonlinear coupling
between the two DoF to observe a time resolved up and down frequency
conversion of the system excitations.

The electronic properties of a symmetric and inductive dc-SQUID at zero-current
bias can be described by a fictitious particle of mass
$m=2C(\Phi_0/2\pi)^2$ moving inside a two-dimensional potential
:
\begin{equation}
  \label{eq:1}
  U(x_{s},x_{a})=2E_J\left[-cosx_scosx_a+\frac{L_J}{L}(x_a-\frac{\pi \Phi_b}{\Phi_0})^2\right]
\end{equation}
    
where $E_J=\Phi_0 I_c/(2\pi)$. Hereafter we will consider the
particle trapped in a local minimum ($x^{min}_{s},x^{min}_{a}$)
verifying $x^{min}_{s}=0$ and $x^{min}_{a}+(L/2L_J)\sin x^{min}_{a}
=\pi \Phi_b/\Phi_0$. By expanding the potential at this minimum up to fourth order,
its quantum dynamics is given by the Hamiltonian
$\mathcal{H}=\mathcal{H}_{s}+\mathcal{H}_{a}+\mathcal{C}_{s,a}$ where $\mathcal{H}_s$ and $\mathcal{H}_a$ are anharmonic
oscillator Hamiltonians describing respectively the symmetric and
antisymmetric plasma modes \cite{Fay_PRB2011,Lecocq_PRL2011}.
$\mathcal{C}_{s,a}$ describes the coupling between the two oscillators.
$\mathcal{H}_{\alpha}=\hbar\omega_p^{\alpha} \left[ (\hat p_{\alpha}^2+ \hat x_{\alpha}^2)/2 -\sigma_{\alpha} \hat x_{\alpha}^3+\delta_{\alpha} \hat x_{\alpha}^4\right]$,
where $\alpha=s,a$ with $({\omega_p^{s}})^2= (2E_J/m)\cos
x^{min}_{a}$ and $({\omega_p^{a}})^2=({\omega_p^{s}})^2+
(4E_J/m)(L_J/L)$. The operators $\hat x_{\alpha}$ and $\hat p_{\alpha}$  are
the reduced position and momentum operators in both directions. The
coupling term has a very simple expression at zero current bias:
\begin{equation}
  \label{eq:2}
  \mathcal{C}_{s,a}=\hbar g_{21}\hat x_{s}^2 \hat x_{a}+\hbar g_{22} \hat x_{s}^2 \hat x_{a}^2
\end{equation}
with $\hbar g_{21}=-E_J(\hbar/m)^\frac{3}{2}(\omega_p^{s}\sqrt{\omega_p^{a}})^{-1}\sin
x^{min}_{a}$ and $\hbar
g_{22}=-E_J(\hbar/m)^{2}(\omega_p^{s}\omega_p^{a})^{-1}\cos
x^{min}_{a}/2$. In the following, we will define
$\ket{n_{s},n_{a}}\equiv \ket{n_{s}}\ket{n_{a}}$ as the eigenstates of
uncoupled hamiltonian $\mathcal{H}_{s}+\mathcal{H}_{a}$, where $\ket{n_{\alpha}}$
indexes the energy levels of each mode. The potentials associated with the
{\it s}-mode and {\it a}-mode are depicted in the figure
Fig.\ref{fig1}b and Fig.\ref{fig1}c for a working point ($I_b=0$,
$\Phi_b=0.37\Phi_0$). At the same bias point the complete spectrum
of the system is presented in Fig.\ref{fig1}d.

The complete aluminum device is fabricated using an angle
evaporation technique without suspended bridges
\cite{Lecocq_Nanotech2011}, and it is presented in Fig.\ref{fig1}a.
The measurements were conducted in a dilution refrigerator at $40$ mK using a standard experimental configuration, previously described in Ref.\cite{Hoskinson_PRL2009}.

\begin{figure}[htb*]
\includegraphics[bb=-120bp 170bp 715bp 670bp,clip,width=0.5\textwidth]{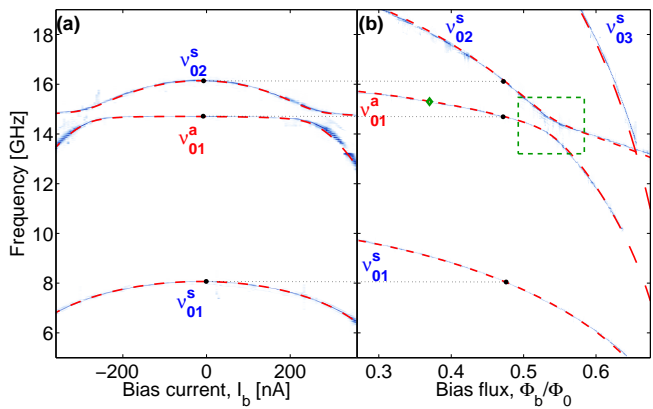}
\caption{\textbf{Energy spectrum}. Escape probability
$P_{esc}$ versus frequency as a function of current bias
\textbf{(a)} and flux bias \textbf{(b)} measured at $\Phi_b
=0.48\Phi_0$ and $I_{b} =0$ respectively. $P_{esc}$ is enhanced when
the frequency matches a resonant transition of the circuit. The
microwave amplitude was tuned to keep the resonance peak amplitude
at 10\%. Dark and bright blue scale correspond to high and small
$P_{esc}$. The red dashed lines are the transition frequencies
deduced from the spectrum of the full hamiltonian with $C=510$~fF
(see text). The green diamond is the initial working point for the measurement of coherent free oscillations between the two modes, presented in Fig.\ref{Oscillations}, and the green dotted square is the area where these oscillations take place.  \label{spectro}}
\end{figure}

The readout of the circuit is performed using switching current techniques. For spectroscopy measurements we apply a microwave pulse field, through the current bias line (see Fig \ref{fig1}a), followed by the readout nanosecond flux pulse that produces a selective escape depending on
the quantum state of the circuit \cite{Claudon_PRL2004,Hoskinson_PRL2009}. The
energy spectrum versus current bias at $\Phi_b=0.48\Phi_0$ and
versus flux bias at $I_b=0$ are plotted in Fig.\ref{spectro}a and
Fig.\ref{spectro}b respectively. In the following we will denote
$\nu^{\alpha}_{nm}$ as the transition frequency between the states
$\ket{n_{\alpha}}$ and $\ket{m_{\alpha}}$, with the other mode in
the ground state. The first transition frequency in
Fig.\ref{spectro} is the one of the camelback phase qubit,
$\nu^{s}_{01}$. With a maximum frequency at zero-current bias, the
system is at an optimal working point with respect to current fluctuations
\cite{Hoskinson_PRL2009}. At higher frequency, the second transition
of the {\it s}-mode is observed with $\nu^{s}_{02}\approx
2\nu^{s}_{01}$. In the flux biased spectrum the third transition
$\nu^{s}_{03}$ is also visible. An additional transition is observed
at about $14.6$~GHz, with a very weak current dependence
(Fig.\ref{spectro}a) but a finite flux dependence
(Fig.\ref{spectro}b). It corresponds to the first transition of the
{\it a}-mode, $\nu^{a}_{01}$. The {\it s}-mode transition
frequencies drop when $\Phi_b/\Phi_0$ approaches $0.7$ which is consistent
with the critical flux $\Phi_c/\Phi_0=1/2+L/(2\pi L_J)= 0.717$ for
which $\omega_p^{s}\rightarrow 0$ and $x^{min}_{a} \rightarrow
\pi/2$. On the contrary $\omega_p^{a}$ remains finite when $\Phi_b
\rightarrow \Phi_c$ with $\omega_p^{a}\rightarrow \sqrt{(4E_J/m)(L_J/L)}$. One
also observes a large level anti-crossing of about $700$~MHz between
the two transitions $\nu^{s}_{02}$ and  $\nu^{a}_{01}$. Additionally
no level anti-crossing is measurable between $\nu^{s}_{03}$ and
$\nu^{a}_{01}$.

The experimental energy spectrum in Fig.\ref{spectro} can be
perfectly fitted by deriving the spectrum of the full Hamiltonian
described above. As all the other parameters of the device can be
extracted from switching-current measurements
\cite{Hoskinson_PRL2009}, the only free parameter is the junction
capacitance. From the fit we obtain $C=510$~fF, which is consistent
with the $10\mu m^2$ junction area. Numerical calculations are used
to describe the energy dependence in Fig.\ref{spectro}a in order to
take into account the current bias and the inductance asymmetry as
well as additional coupling terms. The model describes well the
level anti-crossing between the quantum states $\ket{2_{s},0_{a}}$
and $\ket{0_{s},1_{a}}$, which is given by the nonlinear coupling
term $\hbar g_{21}\hat x_{s}^2\hat x_{a}$. The coupling strength $g_{21}/2\pi$
strongly depends on the working point. It is predicted to change
from zero at $\Phi_b =0$, to $700$~MHz at the anti-crossing,  up to
about $1200$~MHz at $\Phi_b \approx 0.65$ close to the critical line.
The coupling term $\hat x_{s}^2\hat x_{a}$ mainly couples the states
$\ket{0_{s},1_{a}}$ and $\ket{2_{s},0_{a}}$. Starting from these
uncoupled states at $\Phi_b =0$, they become a
maximally entangled state at the resonance condition,
$\nu^{a}_{01}=\nu^{s}_{02}$. Close to $\Phi_b=\Phi_c$ the states are
still entangled because $g_{21}$ diverges. In this device the second
nonlinear coupling term, $\hat x_{s}^2\hat x_{a}^2$, is one order of
magnitude smaller with $g_{22}/2\pi \approx 50$~MHz. Nevertheless the
shift of transition frequencies that it induces must be taken into
account to fit the experimental spectrum. Finally the absence of
$\hat x_{s}^3\hat x_{a}$ in the coupling term explains why no
level anti-crossing between $\nu^{s}_{03}$ and  $\nu^{a}_{01}$ is
observed.

From spectroscopic measurements, we obtain minimum resonance widths
as narrow as $4$~MHz and $3.5~$~MHz for the transitions
$\nu^{s}_{01}$ and $\nu^{a}_{01}$ respectively.
The corresponding dephasing times are estimated to be
 $T_2^s\approx160$~ns and $T_2^a\approx180$~ns. At the
working point ($I_b=0$,$\Phi_b=0.37\Phi_0$) Rabi-like oscillations
are performed on the {\it s}- and {\it a}-mode by applying microwave power
at the resonance frequencies $\nu^{s}_{01}$ and $\nu^{a}_{01}$
respectively. Rabi-decay time is measured in the two-level limit
with about $170$~ns and $50$~ns for {\it s}- and {\it a}-mode.
Relaxation times $T_1^s=200$~ns and $T_1^a=74$~ns are extracted from
the exponential population decay of the excited levels
$\ket{1_{s},0_{a}}$ and $\ket{0_{s},1_{a}}$. The measured
coherence times of the {\it a}-mode are much smaller than expected
from the minimum linewidth. The origin of this additional decoherence
will be discussed later. Nevertheless, the coherence time is
sufficiently large for independent coherent control of each mode.

\begin{figure}[htb*]
\includegraphics[width=0.43\textwidth]{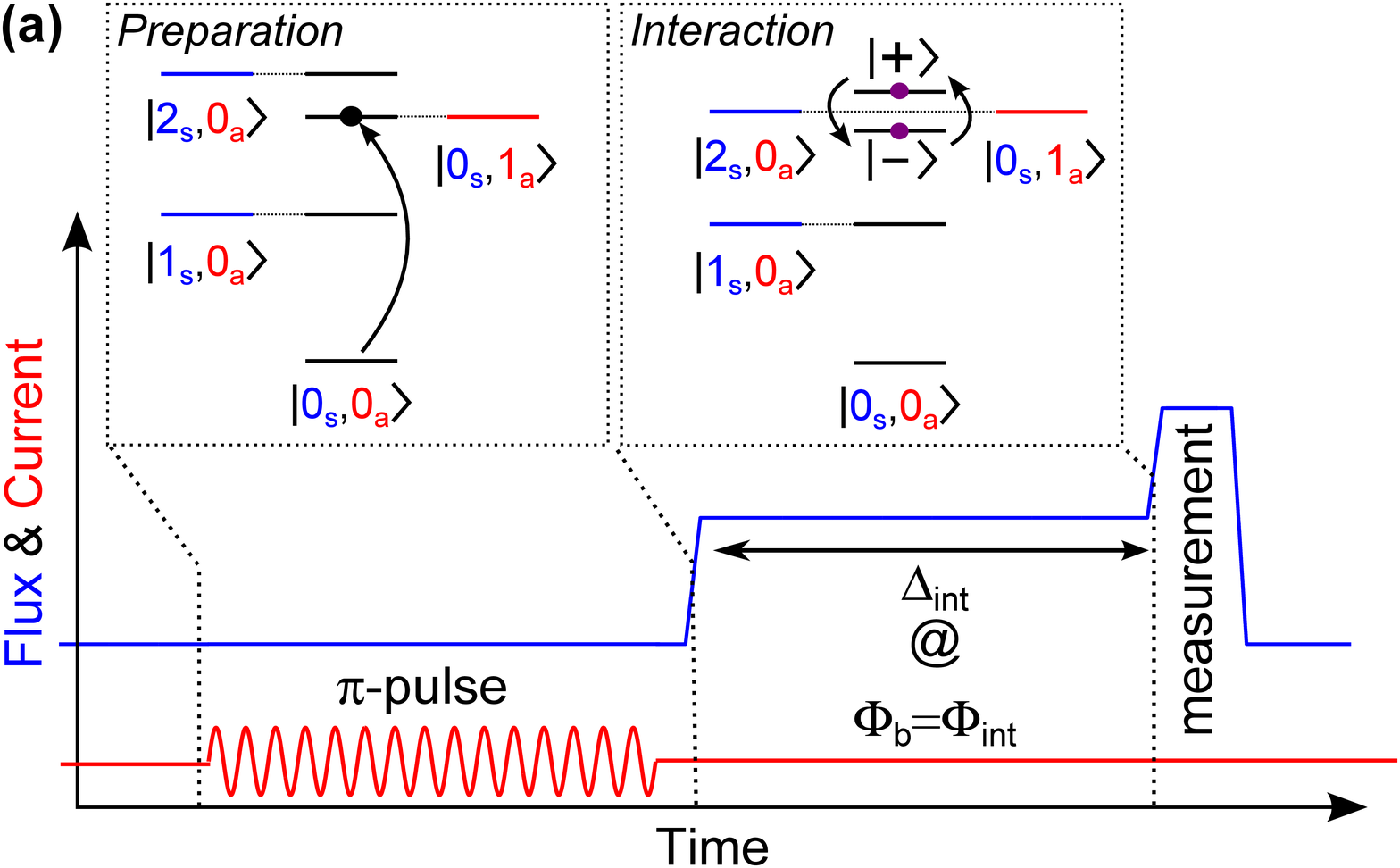}
\includegraphics[bb=-182bp 300bp 778bp 754bp,clip,width=0.49\textwidth]{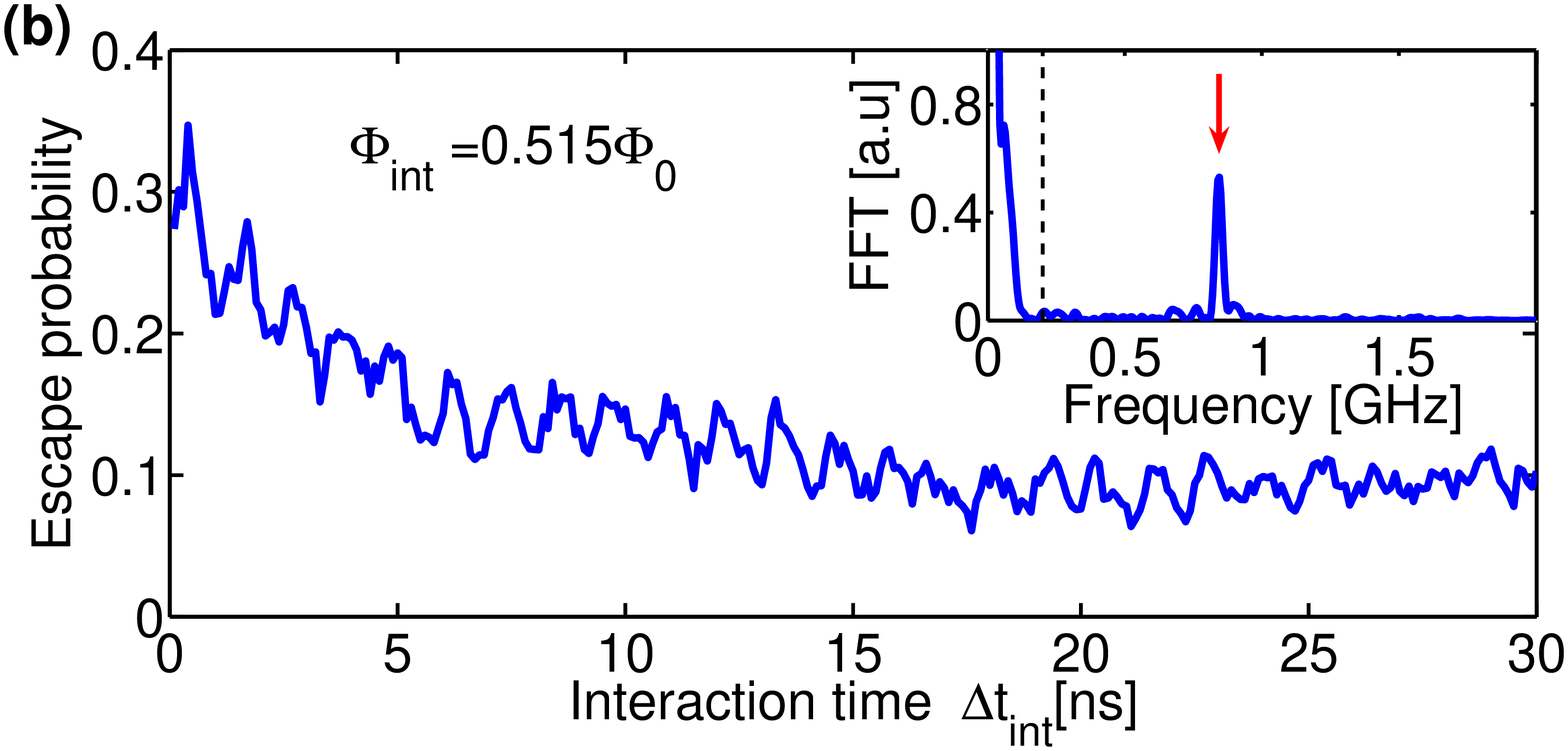}
\includegraphics[bb=-182bp 86bp 778bp 720bp,clip,width=0.49\textwidth]{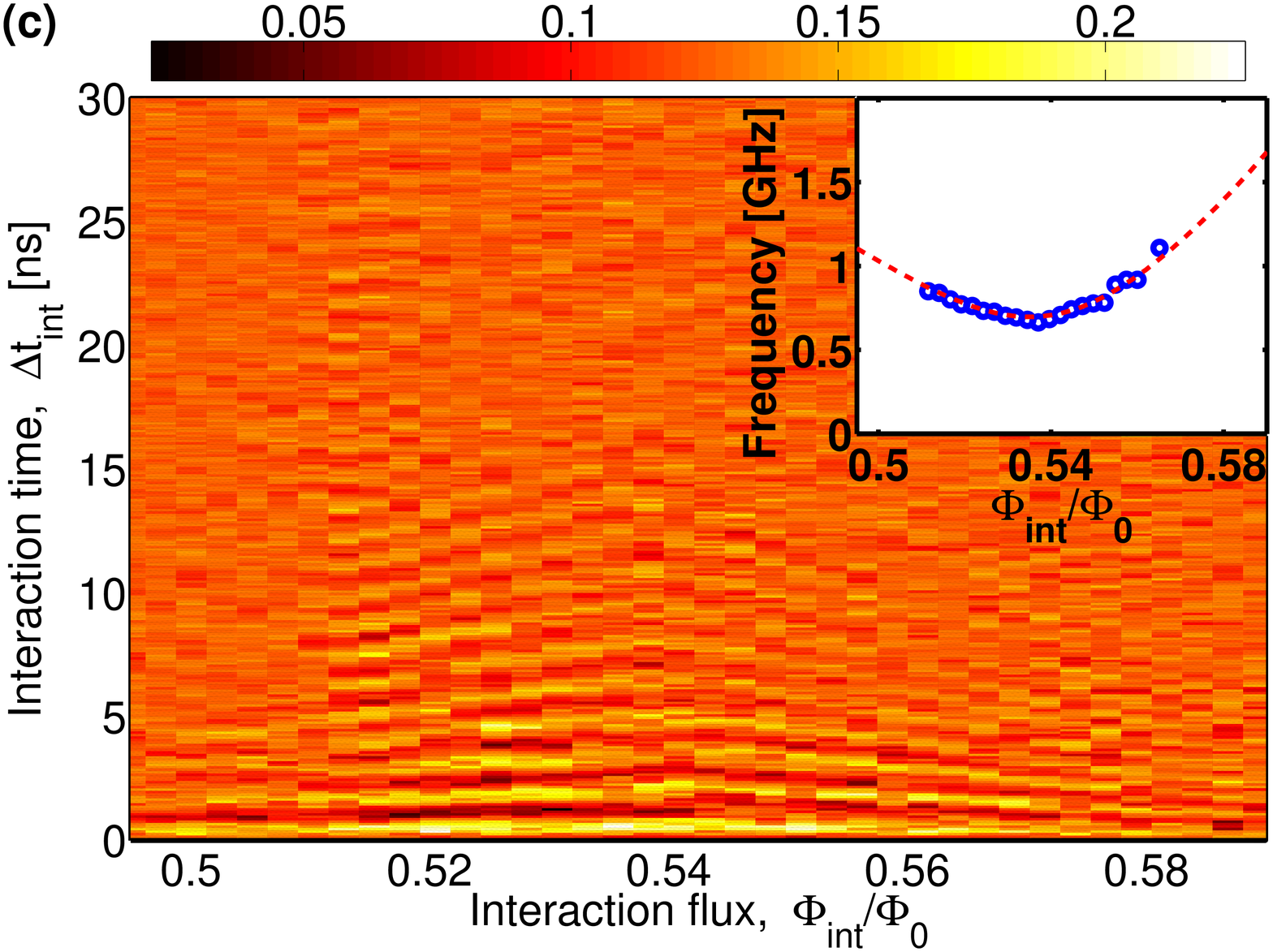}
\caption{ \textbf{Free coherent oscillations between states
$\ket{0_{s},1_{a}}$ and $\ket{2_{s},0_{a}}$ produced by a nonlinear
coupling}. \textbf{(a)} Schematic pulse sequence. The energy diagram, without coupling in blue/red and with coupling in black, is represented for both the preparation and interaction steps.\textbf{(b)} Escape probability $P_{esc}$ versus
interaction time $\Delta t_{int}$. The inset presents the Fourier transform of
these oscillations with a clear peak at $815$~MHz. The red arrow
indicates the theoretically expected frequency. \textbf{(c)}
$P_{esc}$ versus interaction time $\Delta t_{int}$ for different
interaction flux $\Phi_{int}$ close to the resonance condition
between $\nu^s_{02}$ and $\nu^a_{01}$. For clarity the data is
numerically processed using $200$~MHz high-pass filter(dashed line in
inset of \text{(b)}). Inset : oscillation frequency as function of
flux. The dashed red line shows the theoretical predictions.}
\label{Oscillations}
\end{figure}

One of the opportunities given by the rich spectrum of this two DoF
artificial atom is the observation of a coherent
frequency conversion process using the $\hat x_{s}^2\hat x_{a}$ coupling of Eq.\ref{eq:2}. The pulse sequence, similar to other states swapping experiment \cite{McDermott_Science2005,Sillanpaa_Nature2007}, is presented in (see Fig.\ref{Oscillations}a). At
$t=0$, the system is prepared in the state $\ket{0_{s},1_{a}}$, at the
initial working point $\Phi_b =0.37\Phi_0$ (green diamond in Fig.\ref{spectro}b). Immediately after, a non-adiabatic flux pulse \cite{footnote2} brings the system to the working point defined by $\Phi_{int}$,
close to the degeneracy point ($\nu^s_{02} \approx \nu^a_{01}$).
 After the free evolution of the quantum
state during the time $\Delta t_{int}$, we measure the escape probility $P_{esc}$.
Fig.\ref{Oscillations}b presents $P_{esc}$ as a
function of $\Delta t_{int}$ for $\Phi_{int}=0.515\Phi_{0}$. The observed oscillations have a $815MHz$-characteristic frequency (inset of Fig.\ref{Oscillations}b) that matches precisely the theoritical frequency splitting at this flux bias (red arrow).
 In Fig.\ref{Oscillations}c, we present these oscillations as function of $\Phi_{int}$.
 Their frequency varies with $\Phi_{int}$, showing a
typical ``chevron'' pattern. In the inset of
Fig.\ref{Oscillations}c, the oscillation frequency versus
$\Phi_{int}$ is compared to theoretical predictions. The good
agreement between theory and experiment is a striking confirmation
of the observation of swapping between the quantum
states $\ket{0_{s},1_{a}}$ and $\ket{2_{s},0_{a}}$. Instead of the well known
linear coupling $\hat x_{s}\hat x_{a}$ between two oscillators, which
corresponds to a coherent exchange of single excitations between the
two systems, here the coupling $\hat x_{s}^2\hat x_{a}$ is non-linear and
produces a coherent exchange of a single excitation of the {\it
a}-mode with a double excitation of the {\it s}-mode, {\it i.e.} a coherent frequency conversion. Starting from the state $\ket{0_{s},1_{a}}$, an excitation pair $\ket{2_{s},0_{a}}$ is then deterministically produced in about a single nanosecond at the degeneracy point $\Phi_{int}=0.537\Phi_{0}$.

We now discuss the coherent properties and measurement contrast in
our device. The unexpectedly short coherence time of the {\it
a}-mode can be explained by the coupling to spurious two-level
systems (TLS) \cite{Martinis_PRL2005}. With a junction area of $10\mu m^2$ our
device suffers from a large TLS density of about $12$~TLS/GHz (barely
visible in Fig.\ref{spectro}). Therefore it is very difficult to
operate the {\it a}-mode in a frequency window free of TLS since
$\nu^a_{01}$ is only slightly flux dependent. However this is not a
real issue as it can be solved easily by reducing the junction area
\cite{Steffen_PRL2006}. The minimum linewidth of both {\it a}-mode
and {\it s}-mode, and therefore their coherence times, are limited
in our experiment by low frequency flux noise. Operating the system
at $\Phi_b=0$ will lift this limitation since it is an optimal point
with respect to flux noise. The small oscillation amplitude in
Fig.\ref{Oscillations} has two additionnals origins. First the
duration $t_\pi$ of the $\pi$-pulse applied for preparation of the
state $\ket{0_{s},1_{a}}$ has to fulfill the condition
$t_\pi^{-1}<\nu^a_{12}-\nu^a_{01}$ to avoid multilevel dynamics. Since 
$\nu^a_{12}-\nu^a_{01}=18$~MHz in the present experiment,
we fixed $t_\pi=80$~ns. However $t_\pi$ is of the order the relaxation time
 of the {\it a}-mode which implies a strong reduction of the
 $\ket{0_{s},1_{a}}$ occupancy after the $\pi$-pulse, to about $40\%$. The second origin of the
small contrast raises the question of the readout contrast between
the states $\ket{0_{s},1_{a}}$ and $\ket{2_{s},0_{a}}$. In absence
of coupling the escape probability of the state $\ket{0_{s},n_{a}}$
should not be sensitive to $n_{a}$. On the contrary, the escape
probability of the state $\ket{n_{s},0_{a}}$ is very sensitive to
$n_{s}$. This difference should lead to a strong contrast
between the states $\ket{0_{s},1_{a}}$ and $\ket{2_{s},0_{a}}$. However, since the coupling strength
is very large close to the critical lines, $\ket{0_{s},1_{a}}$ and
$\ket{2_{s},0_{a}}$ are still entangled when the escape occurs. This
leads to an additionnal reduction of the contrast between the two states that prevents the extraction of the theoriticaly high efficiency of the frequency conversion \cite{Lecocq_PRL2011}.

In conclusion, we have designed and studied a new type of
superconducting artificial atom with two internal DoF. The spectrum
of this device is well described by considering two anharmonic
oscillators coupled via nonlinear coupling terms. Coherent
manipulation of the two DoF is demonstrated and
the strong nonlinear coupling allows the observation of a coherent frequency conversion between the two internal DoF. If
inserted in a microwave cavity this process should enable parametric
amplification or generation of correlated microwave photons. Finally
by reducing the critical current of the junctions, keeping the ratio
$L_J/L$ constant, one can increase the strength of the coupling term
$g_{22}/2\pi$ up to $500$~~MHz providing a strong dispersive frequency
shift proportional to the population of each mode. This should
allow the realization of C-NOT quantum gate inside
the two DoF of this artificial atom or enable an
efficient readout of the camelback phase qubit state by probing
the transition frequency of the {\it a}-mode.
Considering the long list of unique and interesting features, this new artificial atom with two DoF and a V-shaped energy level structure is a valuable addition to the growing set of building blocks for superconducting quantum electronics.

 We acknowledge C. Hoarau for his help on the electronic setup. We thank P. Milman and M. H. Devoret for fruitful
discussions. We acknowledge the technical support of the PTA facility in CEA Grenoble and of the Nanofab facility in CNRS Grenoble. This work was supported by the European SOLID projects, by the French ANR \textquotedblright QUANTJO\textquotedblright and by the Nanoscience Foundation.


\begin{thebibliography}{99}
\bibitem{footnote2}
With a rise time of 500ps this pulse fulfills the
non-adiabatic condition with respect to the coupling strength
$g_{21}$ ($g_{21}^2/4\approx5$~GHz/ns$<V$ where $V\approx7$~GHz/ns
 is the frequency change velocity in this experiment) but it is still adiabatic with respect to the plasma
frequency of each mode.
\end{thebibliography}
\end{document}